\documentclass[11pt,a4paper]{article}

\usepackage{amssymb}
\usepackage[dvips]{graphicx}
\usepackage{bm}

\begin{document}

\title{\bf The NSVZ scheme for ${\cal N}=1$ SQED with $N_f$ flavors, regularized by the dimensional reduction, in the three-loop approximation}

\author{
S.S.Aleshin, I.O.Goriachuk\\
{\small{\em Moscow State University,}}\\
{\small{\em Faculty of Physics, Department of Theoretical Physics,}}\\
{\small{\em 119991, Moscow, Russia}},\\
\\
A.L.Kataev,\\
{\small{\em Institute for Nuclear Research of the Russian Academy of Science,}}\\
{\small{\em 117312, Moscow, Russia}};\\
{\small{\em Moscow Institute of Physics and Technology,}}\\
{\small{\em 141700, Dolgoprudny, Moscow Region, Russia}},\\
\\
K.V.Stepanyantz\\
{\small{\em Moscow State University,}}\\
{\small{\em Faculty of Physics, Department of Theoretical Physics,}}\\
{\small{\em 119991, Moscow, Russia}}}

\maketitle

\begin{abstract}
At the three-loop level we analyze, how the NSVZ relation appears for ${\cal N}=1$ SQED regularized by the dimensional reduction. This is done by the method analogous to the one which was earlier used for the theories regularized by higher derivatives. Within the dimensional technique, the loop integrals cannot be written as integrals of double total derivatives. However, similar structures can be written in the considered approximation and are taken as a starting point. Then we demonstrate that, unlike the higher derivative regularization, the NSVZ relation is not valid for the renormalization group functions defined in terms of the bare coupling constant. However, for the renormalization group functions defined in terms of the renormalized coupling constant, it is possible to impose boundary conditions to the renormalization constants giving the NSVZ scheme in the three-loop order. They are similar to the all-loop ones defining the NSVZ scheme obtained with the higher derivative regularization, but are more complicated. The NSVZ schemes constructed with the dimensional reduction and with the higher derivative regularization are related by a finite renormalization in the considered approximation.
\end{abstract}

\unitlength=1cm

\vspace*{-20.8cm}

\begin{flushright}
INR-TH-2016-033
\end{flushright}

\vspace*{19.8cm}

\section{Introduction}
\hspace*{\parindent}

Due to non-renormalization theorems  for ${\cal N}=1$ supersymmetric gauge theories, there are no divergent quantum corrections to the superpotential \cite{Grisaru:1979wc} and to the three-point ghost-gauge vertices \cite{Stepanyantz:2016gtk}. Also it is possible to consider as a non-renormalization theorem the so-called NSVZ $\beta$-function \cite{Novikov:1983uc,Jones:1983ip,Novikov:1985rd,Shifman:1986zi}. For ${\cal N}=1$ supersymmetric electrodynamics (SQED) with $N_f$ flavors, which is considered in this paper, it can be written as

\begin{equation}\label{NSVZ_Relation}
\beta(\alpha) = \frac{\alpha^2 N_f}{\pi}\Big(1-\gamma(\alpha)\Big),
\end{equation}

\noindent
where $\gamma(\alpha)$ is the anomalous dimension of the matter superfields \cite{Vainshtein:1986ja,Shifman:1985fi}. It is well-known that Eq. (\ref{NSVZ_Relation}) is scheme-dependent, see, e.g., \cite{Kutasov:2004xu,Kataev:2014gxa}. Therefore, it is necessary to specify the subtraction scheme in which it is obtained. The NSVZ relation is not valid in $\overline{\mbox{DR}}$ \cite{Avdeev:1981ew,Jack:1996vg,Jack:1996cn} and $\mbox{MOM}$ \cite{Kataev:2014gxa,Kataev:2013csa} subtraction schemes. However, it has been explicitly demonstrated \cite{Jack:1996vg,Jack:1996cn,Jack:1998uj} that in the lowest loops one can construct a finite renormalization which relates the $\overline{\mbox{DR}}$ and NSVZ schemes. The possibility of making this finite renormalization is non-trivial, because the terms proportional to $N_f$ in Eq. (\ref{NSVZ_Relation}) are scheme-independent \cite{Kataev:2013csa}.

The NSVZ relation (\ref{NSVZ_Relation}) was derived in all orders by direct summation of supergraphs in \cite{Stepanyantz:2011jy,Stepanyantz:2014ima} for the renormalization group (RG) functions defined in terms of the bare coupling constant in the case of using the supersymmetric version \cite{Krivoshchekov:1978xg,West:1985jx} of the higher derivative regularization \cite{Slavnov:1971aw,Slavnov:1972sq}.\footnote{Note that the RG functions defined in terms of the bare coupling constant are scheme-independent for a fixed regularization.} The RG functions defined in the standard way (in terms of the renormalized coupling constant) coincide with the ones defined in terms of the bare coupling constant, if the boundary conditions

\begin{equation}\label{Boundary_Conditions}
Z_3(\alpha,x_0)=1;\qquad Z(\alpha,x_0)=1
\end{equation}

\noindent
are imposed on the renormalization constants of the charge and of the matter superfields, respectively \cite{Kataev:2013eta}. (Here $x_0$ is a fixed value of $\ln \Lambda_{\mbox{\scriptsize HD}}/\mu$, where $\Lambda_{\mbox{\scriptsize HD}}$ is a dimensionful parameter of the regularized theory, which works as an UV cut-off, and $\mu$ is the normalization point.) Therefore, Eq. (\ref{Boundary_Conditions}) defines the NSVZ scheme in the case of using the higher derivative regularization. This fact has been verified by explicit three-loop calculations \cite{Kataev:2014gxa,Kataev:2013csa,Kataev:2013eta}.

However, at present, supersymmetric theories are mostly regularized by the dimensional reduction \cite{Siegel:1979wq}, which is a special modification of the dimensional regularization \cite{'tHooft:1972fi,Bollini:1972ui,Ashmore:1972uj,Cicuta:1972jf}. It is known that the dimensional reduction is not mathematically consistent \cite{Siegel:1980qs} and breaks supersymmetry in higher loops \cite{Avdeev:1982xy}. Now, there is no prescription similar to Eq. (\ref{Boundary_Conditions}) for theories regularized by the dimensional reduction, and the NSVZ scheme in this case should be constructed by making a specially tuned finite renormalization in each order. In this paper we construct an analog of Eq. (\ref{Boundary_Conditions}) in the three-loop approximation for ${\cal N}=1$ SQED regularized by the dimensional reduction, starting from the structure of the corresponding loop integrals, which was analysed in \cite{Aleshin:2015qqc}.

\section{Three-loop contribution to the $\beta$-function of ${\cal N}=1$ SQED proportional to $(N_f)^2$}
\hspace*{\parindent}\label{Section_Three-Loop}

It is convenient to write the action of ${\cal N}=1$ SQED with $N_f$ flavors in terms of superfields. In the massless limit it has the form

\begin{equation}
S = \frac{1}{4e_0^2} \mbox{Re} \int d^4x\, d^2\theta\, W^a W_a + \frac{1}{4} \sum\limits_{i=1}^{N_f} \int d^4x\, d^4\theta\, \Big(\phi_i^* e^{2V} \phi_i + \widetilde\phi_i^* e^{-2V} \widetilde\phi_i\Big).
\end{equation}

\noindent
In our notation, the two-point Green functions of the gauge superfield and of the matter superfields are written as

\begin{eqnarray}\label{Gamma2}
&& \Gamma^{(2)} - S_{\mbox{\scriptsize gf}} = - \frac{1}{16\pi} \int \frac{d^4p}{(2\pi)^4} d^4\theta\, V(-p,\theta) \partial^2 \Pi_{1/2} V(p,\theta)\, d^{-1}(\alpha_0,\Lambda/p,\varepsilon)
\nonumber\\
&& + \frac{1}{4} \sum\limits_{i=1}^{N_f} \int \frac{d^4p}{(2\pi)^4} d^4\theta \Big(\phi_i^*(-p,\theta) \phi_i(p,\theta) + \widetilde\phi_i^*(-p,\theta) \widetilde \phi_i(p,\theta) \Big) G(\alpha_0,\Lambda/p,\varepsilon),
\end{eqnarray}

\noindent
where $\partial^2\Pi_{1/2} \equiv - D^a\bar D^2 D_a/8$ is the supersymmetric projection operator. The functions $d^{-1}$ and $G$ are calculated by using the dimensional reduction. In our notation, $\varepsilon \equiv 4-d$ and $\Lambda$ is the dimensionful parameter which is introduced to make the coupling constant $e_0$ dimensionless in $d$ dimensions.\footnote{It is important that in our conventions $\Lambda$ does not coincide with the normalization point $\mu$, as it is usually done within the dimensional technique.} Strictly speaking, in Eq. (\ref{Gamma2}) the functions $d^{-1}$ and $G$ should be expressed in terms of the renormalized coupling constant $\alpha$ and the ratio $p/\mu$ in the limit $\varepsilon\to 0$.

In this paper we will start with the result for the function $d^{-1}$ obtained in \cite{Aleshin:2015qqc}, which has been derived by analysing the structure of the loop integrals. It includes the one- and two-loop expressions and also the three-loop terms proportional to $(N_f)^2$:

\begin{eqnarray}\label{Three_Loop_D}
&& d^{-1}(\alpha_0,\Lambda/p,\varepsilon) - \alpha_0^{-1} = 8\pi N_f \Lambda^\varepsilon \int \frac{d^dq}{(2\pi)^d} \frac{1}{q^2(q+p)^2}\nonumber\\
&& - 8\pi N_f \Lambda^{\varepsilon} \frac{\varepsilon}{1-\varepsilon} \int \frac{d^dq}{(2\pi)^d} \frac{1}{q^2(q+p)^2} \left(\ln G(\alpha_0,q/\Lambda,\varepsilon)\right)_{\mbox{\scriptsize 1-loop}}\nonumber\\
&& - 8\pi N_f \Lambda^{\varepsilon} \frac{2\varepsilon}{1-3\varepsilon/2} \int \frac{d^dq}{(2\pi)^d} \frac{1}{q^2(q+p)^2} \left(\ln G(\alpha_0,q/\Lambda,\varepsilon)\right)_{\mbox{\scriptsize 2-loop}, N_f}\nonumber\\
&& + \mbox{finite terms} + O(\alpha_0^2 N_f) + O(\alpha_0^3),\vphantom{\frac{d^dk}{(2\pi)^d}}
\end{eqnarray}

\noindent
where $\ln G_{\mbox{\scriptsize 1-loop}}$ is a part of the function $\ln G$ proportional to $\alpha_0$, and $(\ln G)_{\mbox{\scriptsize 2-loop}, N_f}$ is a part of $\ln G$ proportional to $\alpha_0^2 N_f$.

The expression (\ref{Three_Loop_D}) is the analog of a simple relation between the two-point Green functions of the gauge superfield and of the matter superfields which is valid in all orders in the case of using the higher derivative regularization \cite{Stepanyantz:2011jy,Stepanyantz:2014ima}. In the case of the higher derivative regularization it follows from the factorization of the integrals defining the $\beta$-function into integrals of (double) total derivatives, which was noted in \cite{Soloshenko:2003nc,Smilga:2004zr} and has been confirmed by numerous explicit calculations in various supersymmetric theories \cite{Pimenov:2009hv,Stepanyantz:2011bz,Kazantsev:2014yna,Aleshin:2016yvj}. With the dimensional reduction this property does not take place, because the limit of the vanishing external momentum is not well-defined. That is why the three-loop result (\ref{Three_Loop_D}) is more complicated and does not lead to the NSVZ $\beta$-function for the RG functions defined in terms of the bare coupling constant.

Let us, following Ref. \cite{Chetyrkin:1980pr}, introduce the notation

\begin{equation}
G(\alpha,\beta) \equiv \frac{\Gamma(\alpha+\beta-2+\varepsilon/2)}{\Gamma(\alpha)\Gamma(\beta)} B(2-\alpha-\varepsilon/2,2-\beta-\varepsilon/2).
\end{equation}

\noindent
Then, if the dimensional reduction is used for regularization, the function $\ln G$ can be presented in the form\footnote{Although the dimensional reduction can lead to some potential problems in higher orders, in the considered approximation it works properly.}

\begin{equation}\label{Logarithm}
\ln G = \sum\limits_{n=1}^\infty (\alpha_0)^n \Big(\frac{4\pi\Lambda^2}{q^2}\Big)^{\varepsilon n/2} g_n(\varepsilon),
\end{equation}

\noindent
where only the coefficients

\begin{eqnarray}\label{G1}
&& g_1(\varepsilon) = - \frac{1}{2\pi}\,G(1,1);\\
\label{G2}
&& g_2(\varepsilon) = \frac{1}{4\pi^2}\, (1+N_f) G(1,1) G(1,1+\varepsilon/2) - \frac{1}{8\pi^2}\, G(1,1)^2 + \mbox{finite terms}\qquad
\end{eqnarray}

\noindent
are essential in the two-loop approximation.

\section{RG functions defined in terms of the bare coupling constant}
\hspace{\parindent}\label{Section_Bare_RG}

In the case of using the higher derivative regularization the NSVZ equation relates the RG functions defined in terms of the bare coupling constant. The expression (\ref{Three_Loop_D}) is constructed in the way analogous to the derivation of the NSVZ $\beta$-function with the higher derivative regularization. That is why in this section we try to derive the NSVZ-like relation for the RG functions defined in terms of the bare coupling constant from Eq. (\ref{Three_Loop_D}). These functions are defined according to the following prescription:

\begin{eqnarray}\label{Bare_RG_Functions}
\beta(\alpha_0)\equiv \frac{d\alpha_0}{d\ln\Lambda}\Big|_{\alpha=\mbox{\scriptsize const}};\qquad \gamma(\alpha_0) \equiv  - \frac{d\ln Z}{d\ln\Lambda}\Big|_{\alpha=\mbox{\scriptsize const}},
\end{eqnarray}

\noindent
where $\alpha = e^2/4\pi = \alpha(\alpha_0,\varepsilon,\Lambda/\mu)$ is the renormalized coupling constant. It is known \cite{Kataev:2013eta} that the RG functions (\ref{Bare_RG_Functions}) depend on regularization, but are scheme independent if a regularization is fixed. The differentiation in Eq. (\ref{Bare_RG_Functions}) should be made at a fixed value of the renormalized coupling constant. That is why it is necessary to express the bare coupling constant in terms of the renormalized one by using the equation

\begin{equation}
\alpha_0 = \alpha + \frac{\alpha^2 N_f}{\pi} \Big(\frac{1}{\varepsilon} + \ln \frac{\bar\Lambda}{\mu} + b_1\Big) + O(\alpha^3),
\end{equation}

\noindent
where $\bar\Lambda \equiv \Lambda \exp(-\gamma/2) \sqrt{4\pi}$. Note that this expression also contains an arbitrary constant $b_1$ which defines the subtraction scheme in the considered approximation.

Taking into account that the expression $ZG$ is finite by construction, it is possible to relate the anomalous dimension (\ref{Bare_RG_Functions}) to the derivative of the Green function $G$ with respect to $\ln\Lambda$,

\begin{eqnarray}\label{Anomalous_Dimension}
&& \gamma(\alpha_0) = \lim\limits_{\varepsilon\to 0}\, \frac{d}{d\ln\Lambda} \ln G\Big(\alpha_0(\alpha,\Lambda/\mu,\varepsilon),\Lambda/q,\varepsilon\Big)\Big|_{\alpha_0 = \alpha_0(\alpha,\varepsilon,\Lambda/\mu);\ \alpha=\mbox{\scriptsize const}}\nonumber\\
&& = \lim\limits_{\varepsilon\to 0}\, \Bigg(\Big(\frac{4\pi\Lambda^2}{q^2}\Big)^{\varepsilon/2} g_1(\varepsilon) \Big(\varepsilon \alpha_0 + \frac{\alpha_0^2 N_f}{\pi} \Big) + 2\varepsilon \alpha_0^2 \Big(\frac{4\pi\Lambda^2}{q^2}\Big)^{\varepsilon} g_2(\varepsilon)\Bigg)\Bigg|_{\alpha_0=\alpha_0(\alpha,\varepsilon,\Lambda/\mu)}\nonumber\\
&& + O(\alpha_0^3).\vphantom{\frac{1}{2}}
\end{eqnarray}

\noindent
It is important that to calculate this expression, first, one should express the bare coupling constant in terms of the renormalized one. Next, it is necessary to remove all terms vanishing in the limit $\varepsilon \to 0$\footnote{Although we denote this operation by $\lim\limits_{\varepsilon\to 0}$, strictly speaking, the limit does not exist due to $\varepsilon$-poles, which should be kept.} and, finally, express the result in terms of the bare coupling constant. After these transformations we obtain

\begin{equation}\label{Bare_Two_Loop_Gamma}
\gamma(\alpha_0) = - \frac{\alpha_0}{\pi} + \frac{\alpha_0^2}{2\pi^2} (1+ N_f) + O(\alpha_0^3).
\end{equation}

\noindent
One can note that this anomalous dimension coincides with the one defined in terms of the renormalized coupling constant in the $\overline{\mbox{DR}}$-scheme. In the next section we will demonstrate that it is not an accident. However, now, we proceed  to calculating the $\beta$-function defined in terms of the bare coupling constant by Eq. (\ref{Bare_RG_Functions}). This can be done using the equation

\begin{equation}
\frac{\beta(\alpha_0)}{\alpha_0^2} = \lim\limits_{\varepsilon\to 0}\frac{d}{d\ln\Lambda}\Big(d^{-1}(\alpha_0,\Lambda/p,\varepsilon) - \alpha_0^{-1}\Big)\Big|_{\alpha=\mbox{\scriptsize const}},
\end{equation}

\noindent
which follows from the fact that $d^{-1}$ is a finite function of the renormalized coupling constant $\alpha$. Again, in calculating this expression it is necessary to express $\alpha_0$ in terms of $\alpha$ and, after this, remove the terms vanishing in the limit $\varepsilon\to 0$. The result should depend on $\alpha_0$ and does not contain logarithms.

Substituting Eq. (\ref{Logarithm}) into Eq. (\ref{Three_Loop_D}), calculating the integrals, and differentiating the result with respect to $\ln\Lambda$ (taking into account that $\alpha_0$ also depends on $\Lambda$) we obtain

\begin{eqnarray}\label{Beta_Expression}
&&\hspace*{-5mm} \frac{\beta(\alpha_0)}{\alpha_0^2} = \lim\limits_{\varepsilon\to 0}\Bigg(\frac{N_f\varepsilon}{2\pi} \Big(\frac{4\pi\Lambda^2}{p^2}\Big)^{\varepsilon/2} G(1,1) - \frac{N_f}{\pi} \frac{\varepsilon}{1-\varepsilon} \Big(\frac{4\pi\Lambda^2}{p^2}\Big)^{\varepsilon} G(1,1+\varepsilon/2)\, g_1(\varepsilon) \Big(\varepsilon \alpha_0 + \frac{\alpha_0^2 N_f}{2\pi}\Big) \nonumber\\
&&\hspace*{-5mm} - \frac{\alpha_0^2 N_f}{\pi} \frac{3\varepsilon^2}{1-3\varepsilon/2} \Big(\frac{4\pi\Lambda^2}{p^2}\Big)^{3\varepsilon/2} G(1,1+\varepsilon)\, g_2(\varepsilon)\Bigg)\Bigg|_{\alpha_0=\alpha_0(\alpha,\varepsilon,\Lambda/\mu)} + O(\alpha_0^2 N_f) + O(\alpha_0^3).\vphantom{\frac{d^dk}{(2\pi)^d}}
\end{eqnarray}

\noindent
(Note that only a part of the function $g_2(\varepsilon)$ proportional to $N_f$ is essential in the considered approximation.) The limit $\varepsilon\to 0$ in the one-loop contribution can be taken straightforwardly. However, in calculating the next terms, it is necessary to take into account that

\begin{equation}\label{G_Asymptotics}
g_1(\varepsilon) =-\frac{1}{\pi\varepsilon} + O(1);\qquad g_2(\varepsilon) = \frac{N_f}{2\pi^2\varepsilon^2} + O(\varepsilon^{-1}).
\end{equation}

\noindent
Then it is convenient to rewrite Eq. (\ref{Beta_Expression}) in the form

\begin{eqnarray}
&&\hspace*{-5mm} \frac{\beta(\alpha_0)}{\alpha_0^2} = \frac{N_f}{\pi}  + \lim\limits_{\varepsilon\to 0}\Bigg( - \frac{N_f}{\pi}\, \Big(\frac{4\pi\Lambda^2}{p^2}\Big)^{\varepsilon/2} g_1(\varepsilon) \Big(\varepsilon \alpha_0 + \frac{\alpha_0^2 N_f}{\pi}\Big) -\frac{2N_f \varepsilon\alpha_0^2}{\pi} \Big(\frac{4\pi\Lambda^2}{p^2}\Big)^{\varepsilon} g_2(\varepsilon)\Bigg)\nonumber\\
&&\hspace*{-5mm} - \frac{N_f}{\pi}\cdot \lim\limits_{\varepsilon\to 0} \Bigg(\Big(\frac{4\pi\Lambda^2}{p^2}\Big)^{\varepsilon/2}\Big[\frac{\varepsilon}{1-\varepsilon} \Big(\frac{4\pi\Lambda^2}{p^2}\Big)^{\varepsilon/2} G(1,1+\varepsilon/2) -1\Big]\,g_1(\varepsilon)  \Big(\varepsilon \alpha_0 + \frac{\alpha_0^2 N_f}{\pi}\Big) - \frac{\varepsilon}{1-\varepsilon} \nonumber\\
&&\hspace*{-5mm} \times \Big(\frac{4\pi\Lambda^2}{p^2}\Big)^{\varepsilon} G(1,1+\varepsilon/2)\,g_1(\varepsilon)\, \frac{\alpha_0^2 N_f}{2\pi} + 2 \alpha_0^2
\Big(\frac{4\pi\Lambda^2}{p^2}\Big)^{\varepsilon} \Big[\frac{3\varepsilon/2}{1-3\varepsilon/2} \Big(\frac{4\pi\Lambda^2}{p^2}\Big)^{\varepsilon/2} G(1,1+\varepsilon) -1 \Big]\nonumber\\
&&\hspace*{-5mm} \times \varepsilon g_2(\varepsilon)\Bigg)\Bigg|_{\alpha_0=\alpha_0(\alpha,\varepsilon,\Lambda/\mu)} + O(\alpha_0^2 N_f) + O(\alpha_0^3).\vphantom{\Bigg|}
\end{eqnarray}

\noindent
Comparing this expression with Eq. (\ref{Anomalous_Dimension}) we see that the first line gives the NSVZ relation in the considered approximation for the RG functions defined in terms of the bare coupling constant. Then taking the identity

\begin{equation}
\lim\limits_{\varepsilon \to 0} (\varepsilon \alpha_0) = \lim\limits_{\varepsilon \to 0} \varepsilon \Big(\alpha + \frac{\alpha^2 N_f}{\pi} \Big(\frac{1}{\varepsilon} + \ln \frac{\bar\Lambda}{\mu} + b_1\Big) + O(\alpha^3)\Big) = \frac{\alpha^2 N_f}{\pi} + O(\alpha^3) = \frac{\alpha_0^2 N_f}{\pi} + O(\alpha_0^3)
\end{equation}

\noindent
into account and using Eqs. (\ref{G_Asymptotics}) and (\ref{G1}), it is possible to present the expression under consideration in the form

\begin{equation}
\frac{\beta(\alpha_0)}{\alpha_0^2} = \frac{N_f}{\pi}\Big(1 - \gamma(\alpha_0) \Big) + \frac{\Delta\beta(\alpha_0)}{\alpha_0^2}
+ O(\alpha_0^2 N_f) + O(\alpha_0^3),
\end{equation}

\noindent
where we introduce the notation

\begin{eqnarray}
&& \Delta\beta(\alpha_0) = \frac{\alpha_0^4 (N_f)^2}{\pi^3}\cdot \lim\limits_{\varepsilon\to 0}
\Bigg(\frac{2}{\varepsilon}\Big[\frac{\varepsilon}{1-\varepsilon} \Big(\frac{4\pi\Lambda^2}{p^2}\Big)^{\varepsilon/2} G(1,1+\varepsilon/2) - 1\Big]  - \frac{\varepsilon}{4(1-\varepsilon)}\qquad\nonumber\\
&&\times \Big(\frac{4\pi\Lambda^2}{p^2}\Big)^{\varepsilon} G(1,1) G(1,1+\varepsilon/2)  - \frac{1}{\varepsilon} \Big[\frac{3\varepsilon/2}{1-3\varepsilon/2} \Big(\frac{4\pi \Lambda^2}{p^2}\Big)^{\varepsilon/2} G(1,1+\varepsilon) -1\Big]\Bigg).
\end{eqnarray}

The (scheme-independent) expression $\Delta\beta(\alpha_0)$ can be simplified in the $G$-scheme \cite{Chetyrkin:1980pr} (see also \cite{Kataev:1988sq}). Let us introduce the parameter $\Lambda_G$ defined by the equation

\begin{equation}
(\Lambda_G)^\varepsilon \equiv \frac{\varepsilon}{2} G(1,1) (\sqrt{4\pi}\Lambda)^\varepsilon
\end{equation}

\noindent
instead of $\Lambda$. Then, after some transformations involving the well-known identities

\begin{equation}
\Gamma(1+x) = x \Gamma(x);\qquad B(x,y) = \frac{\Gamma(x)\Gamma(y)}{\Gamma(x+y)},
\end{equation}

\noindent
the expression for $\Delta\beta$ can be rewritten in the form

\begin{eqnarray}
&& \Delta\beta(\alpha_0) = \frac{\alpha_0^4 (N_f)^2}{\pi^3}\cdot \lim\limits_{\varepsilon\to 0}
\Bigg(\frac{2}{\varepsilon(1-\varepsilon^2)} \Big(\frac{\Lambda_G}{p}\Big)^{\varepsilon} \frac{B(1-\varepsilon/2,1-\varepsilon)}{B(1+\varepsilon/2,1+\varepsilon/2) B(1-\varepsilon/2,1-\varepsilon/2)}\nonumber\\
&& \times \Big(1-\frac{1}{4} \Big(\frac{\Lambda_G}{p}\Big)^{\varepsilon}\Big)- \frac{1}{\varepsilon(1-9\varepsilon^2/4)} \Big(\frac{\Lambda_G}{p}\Big)^{\varepsilon} \frac{B(1-\varepsilon/2,1-3\varepsilon/2)}{B(1+\varepsilon/2, 1+\varepsilon) B(1-\varepsilon/2,1-\varepsilon/2)} -\frac{1}{\varepsilon}\Bigg).\qquad
\end{eqnarray}

\noindent
The limit in this expression can be easily taken using the equation

\begin{equation}
B(1+x\varepsilon, 1+ y\varepsilon) = 1 - (x+y)\varepsilon + O(\varepsilon^2).
\end{equation}

\noindent
Then, after some transformations, we obtain

\begin{equation}
\Delta\beta = \frac{\alpha_0^4 (N_f)^2}{\pi^3} \Big(-\frac{1}{2\varepsilon} -\frac{1}{4}\Big).
\end{equation}

\noindent
This implies that the result for the $\beta$-function defined in terms of the bare coupling constants in the case of using the dimension reduction can be written in the form

\begin{equation}\label{Bare_Three-Loop_Beta}
\frac{\beta(\alpha_0)}{\alpha_0^2} = \frac{N_f}{\pi}\Big(1-\gamma(\alpha_0)\Big) + \frac{\alpha_0^2 (N_f)^2}{\pi^3}\Big(-\frac{1}{2\varepsilon} -\frac{1}{4}\Big) + O(\alpha_0^2 N_f) + O(\alpha_0^3).
\end{equation}

\noindent
From this equation we see that, unlike the higher derivative regularization, the RG functions defined in terms of the bare coupling constant do not satisfy the NSVZ relation. Moreover, we see that the $\beta$-function explicitly depends on $\varepsilon$. However, in the next section we will demonstrate that the $\beta$-function defined in terms of the renormalized coupling constant is $\varepsilon$-independent and, in the $\overline{\mbox{DR}}$-scheme, is related to the $\beta$-function (\ref{Bare_RG_Functions}). Moreover, we will see that $-1/4$ in the second term determines the finite renormalization which relates the $\overline{\mbox{DR}}$ and NSVZ schemes.

\section{$\overline{\mbox{DR}}$-scheme}
\hspace*{\parindent}\label{Section_DR}

In the previous section we deal with the RG functions defined in terms of the bare coupling constant. However, standardly, the RG functions are defined in terms of the renormalized coupling constant,

\begin{eqnarray}\label{Renormalized_RG_Functions}
\widetilde\beta(\alpha)\equiv \frac{d\alpha}{d\ln\mu}\Big|_{\alpha_0=\mbox{\scriptsize const}};\qquad \widetilde\gamma(\alpha) \equiv  \frac{d\ln Z}{d\ln\mu}\Big|_{\alpha_0=\mbox{\scriptsize const}},
\end{eqnarray}

\noindent
where $\mu$ is the normalization point, which is an argument of the renormalized quantities. It is well known that these functions are scheme dependent. For the theory  regularized by higher derivatives the RG functions defined in terms of the bare coupling constant and the RG functions defined in terms of the renormalized coupling constant coincide ($\widetilde\beta(\alpha)=\beta(\alpha)$; $\widetilde\gamma(\alpha)= \gamma(\alpha)$), if the boundary conditions

\begin{equation}\label{NSVZ_HD_Scheme}
\alpha_0(\alpha,x_0)=\alpha;\qquad Z(\alpha,x_0)=1,
\end{equation}

\noindent
are imposed on the renormalization constants \cite{Kataev:2014gxa,Kataev:2013csa,Kataev:2013eta}, where $x_0$ is a fixed value of $\ln\Lambda_{\mbox{\scriptsize HD}}/\mu$. The boundary conditions (\ref{NSVZ_HD_Scheme}) (which can be equivalently presented in the form (\ref{Boundary_Conditions})) define the NSVZ scheme for the considered theory with the higher derivative regularization in all orders.

Now, let us find analogs of these constructions for the theory regularized by the dimensional reduction. There are two main differences:

1. The RG functions defined in terms of the bare coupling constant do not satisfy the NSVZ relation, see Eq. (\ref{Bare_Three-Loop_Beta}).

2. The renormalization constants depend not only on $\ln\Lambda/\mu$, but also on $\varepsilon$.

Note that usually with the dimensional reduction one sets $\Lambda=\mu$. However, we do not impose this condition to make the calculations similar to the case of the higher derivative regularization. That is why to construct the $\overline{\mbox{DR}}$-scheme, we include only $\varepsilon$-poles and powers of $\ln\bar\Lambda/\mu$ into the renormalization constants.

Let us formally consider the functions $\alpha_0(\alpha,\varepsilon, x)$ and $Z(\alpha,\varepsilon, x)$, where $x\equiv \ln \bar\Lambda/\mu$. For example, for ${\cal N}=1$ SQED investigated in this paper from Eqs. (\ref{Bare_Three-Loop_Beta}) and (\ref{Bare_Two_Loop_Gamma}) one obtains

\begin{eqnarray}\label{Three-Loop_Alpha}
&& \frac{1}{\alpha_0(\alpha,\varepsilon,x)} = \frac{1}{\alpha} - \frac{N_f}{\pi}\Big(\frac{1}{\varepsilon} + \ln\frac{\bar\Lambda}{\mu} + b_1\Big)
-\frac{\alpha N_f}{\pi^2} \Big(\frac{1}{2\varepsilon} + \ln\frac{\bar\Lambda}{\mu} + b_2\Big)  - \frac{\alpha^2 N_f^2}{\pi^3} \Big(\frac{1}{6\varepsilon^2}\nonumber\\
&& -\frac{1}{4\varepsilon} + \frac{1}{2\varepsilon}\ln\frac{\bar\Lambda}{\mu} + \frac{b_1}{2\varepsilon} + \frac{1}{2} \ln^2 \frac{\bar\Lambda}{\mu} +b_1 \ln\frac{\bar\Lambda}{\mu} -\frac{3}{4}\ln\frac{\bar\Lambda}{\mu} + b_3\Big)
+ O(\alpha^2 N_f) + O(\alpha^3);\qquad\\
\label{Two-Loop_Z}
&& \ln Z(\alpha,\varepsilon,x) = \frac{\alpha}{\pi}\Big(\frac{1}{\varepsilon} + \ln\frac{\bar\Lambda}{\mu} + g_1\Big) + \frac{\alpha^2}{\pi^2}\Big(-\frac{1}{4\varepsilon} - \frac{1}{2}\ln\frac{\bar\Lambda}{\mu} + g_2 -\frac{g_1^2}{2} \Big) + \frac{\alpha^2 N_f}{\pi^2}\nonumber\\
&& \times \Big(\frac{1}{2\varepsilon^2} + \frac{1}{\varepsilon}\ln\frac{\bar\Lambda}{\mu} +\frac{1}{2}\ln^2\frac{\bar\Lambda}{\mu} -\frac{1}{4\varepsilon} -\frac{1}{2}\ln\frac{\bar\Lambda}{\mu} + \frac{b_1}{\varepsilon} + b_1 \ln \frac{\bar\Lambda}{\mu}\Big) + O(\alpha^3)
\end{eqnarray}

\noindent
by integrating the RG equations (\ref{Bare_RG_Functions}). Note that to find the dependence on $\varepsilon$, it is necessary to take into account that in $L$-loops terms linear in $\varepsilon$ and $\ln\bar\Lambda/\mu$ in the $\overline{\mbox{DR}}$-scheme are present in the combination

\begin{equation}
\frac{1}{L\varepsilon} + \ln \frac{\bar\Lambda}{\mu}.
\end{equation}

\noindent
Higher order $\varepsilon$-poles can be found using the standard renormalization group technique \cite{'tHooft:1973mm} (see also the lectures \cite{Kazakov:2008tr}). However, the results (\ref{Three-Loop_Alpha}) and (\ref{Two-Loop_Z}) are not uniquely defined due to the arbitrariness of choosing the subtraction scheme. Therefore, these equations contain arbitrary finite constants $b_i$ and $g_i$. To specify the subtraction scheme one should fix values of these constants by a special additional prescription.

Let us formally take the limit $\varepsilon \to \infty$ of Eqs. (\ref{Three-Loop_Alpha}) and (\ref{Two-Loop_Z}). In this limit the renormalization constants depend only on $x = \ln\bar\Lambda/\mu$, as in the case of the higher derivative regularization. By construction, the $\overline{\mbox{DR}}$-scheme is defined by the equations

\begin{equation}\label{DR_Boundary_Conditions}
\alpha_0(\alpha,\varepsilon\to\infty, x=0) = \alpha;\qquad Z(\alpha,\varepsilon\to\infty, x=0)=1.
\end{equation}

\noindent
(The first equality can be also rewritten as $Z_3(\alpha,\varepsilon\to\infty, x=0)=1$.) Eq. (\ref{DR_Boundary_Conditions}) is analogous to Eq. (\ref{NSVZ_HD_Scheme}) with $x_0=0$ and coincides with it after the substitution $\Lambda_{\mbox{\scriptsize HD}} \to \bar\Lambda$. Therefore, repeating the argumentation of \cite{Kataev:2013eta} we find

\begin{equation}\label{RG_Relation}
\widetilde\beta_{\overline{\mbox{\scriptsize DR}}}(\alpha) = \lim\limits_{\varepsilon\to \infty}\beta(\alpha_0,\varepsilon)\Big|_{\alpha_0=\alpha};\qquad \widetilde\gamma_{\overline{\mbox{\scriptsize DR}}}(\alpha) = \lim\limits_{\varepsilon\to \infty}\gamma(\alpha_0,\varepsilon)\Big|_{\alpha_0=\alpha}.
\end{equation}

\noindent
This implies that after eliminating $\varepsilon$-poles in the RG functions defined in terms of the bare coupling constant, they coincide with the RG functions in $\overline{\mbox{DR}}$-scheme.

To verify this statement, we consider the above example. After some transformations, from Eqs. (\ref{Three-Loop_Alpha}) and (\ref{Two-Loop_Z}) we obtain

\begin{eqnarray}
&& \widetilde\beta_{\overline{\mbox{\scriptsize DR}}}(\alpha) = \frac{\alpha^2 N_f}{\pi}\Big(1+\frac{\alpha}{\pi} - \frac{3 \alpha^2 N_f}{4\pi^2}\Big) + O(\alpha^4 N_f) + O(\alpha^5)= \lim\limits_{\varepsilon\to \infty}\beta(\alpha_0,\varepsilon)\Big|_{\alpha_0=\alpha};\qquad\\
&& \widetilde\gamma_{\overline{\mbox{\scriptsize DR}}}(\alpha) =  - \frac{\alpha}{\pi} + \frac{\alpha^2}{2\pi^2} (1+ N_f) + O(\alpha^3) = \lim\limits_{\varepsilon\to \infty}\gamma(\alpha_0,\varepsilon)\Big|_{\alpha_0=\alpha}.
\end{eqnarray}

\noindent
Thus, we have verified that Eq. (\ref{RG_Relation}) is really valid in this case. Comparing the expressions for the $\beta$-function and for the anomalous dimension of the matter superfields in the $\overline{\mbox{DR}}$-scheme we see that the equation analogous to the NSVZ relation can be written as

\begin{equation}\label{Modified_NSVZ}
\frac{\widetilde\beta_{\overline{\mbox{\scriptsize DR}}}(\alpha)}{\alpha^2} = \frac{N_f}{\pi}\Big(1-\gamma_{\overline{\mbox{\scriptsize DR}}}(\alpha)\Big) - \frac{\alpha^2 (N_f)^2}{4\pi^3} + O(\alpha^3).
\end{equation}

\noindent
This expression is in agreement with the result of Ref. \cite{Jack:1996vg}, where the $\beta$-function and the anomalous dimension were calculated separately.
Note that in writing Eq. (\ref{Modified_NSVZ}) we take into account that, in the Abelian case, the terms linear in $N_f$ are scheme-independent in all orders \cite{Kataev:2013vua} and always satisfy the NSVZ relation \cite{Kataev:2013csa}. We see that Eq. (\ref{Modified_NSVZ}) coincides with the limit $\varepsilon\to \infty$ of Eq. (\ref{Bare_Three-Loop_Beta}), in agreement with the above argumentation.

\section{NSVZ scheme in the three-loop approximation}
\hspace*{\parindent}\label{Section_Scheme}

Now, let us construct the boundary conditions analogous to Eq. (\ref{NSVZ_HD_Scheme}) in the case of using the dimensional reduction in the considered approximation. It is well known that the NSVZ-scheme and $\overline{\mbox{DR}}$-scheme can be related by the finite renormalization

\begin{equation}\label{Finite_Renormalization}
\alpha'=\alpha'(\alpha);\qquad \phi_R' = z(\alpha)^{-1/2} \phi_R,
\end{equation}

\noindent
where $\alpha'(\alpha)$ and $z(\alpha)$ are finite functions \cite{Jack:1996vg,Jack:1996cn,Jack:1998uj}. Under this finite renormalization the RG functions are changed as

\begin{equation}\label{RG_Transformation}
\widetilde\beta'(\alpha') = \frac{d\alpha'}{d\alpha}\cdot \widetilde\beta(\alpha);\qquad \widetilde\gamma'(\alpha')= \frac{d\ln z}{d\alpha}\cdot \widetilde\beta(\alpha) + \widetilde\gamma(\alpha).
\end{equation}

\noindent
Let us assume that $\alpha = \alpha_{\overline{\mbox{\scriptsize DR}}}$ and $\alpha'=\alpha_{\mbox{\scriptsize NSVZ}}$ and set $z(\alpha)=1$. Then from the above equations one can obtain \cite{Kutasov:2004xu,Kataev:2013csa}

\begin{equation}
\widetilde\beta(\alpha) = \frac{d\alpha}{d\alpha'} \cdot \frac{\alpha'{}^2 N_f}{\pi} \Big(1-\widetilde\gamma(\alpha)\Big)\Big|_{\alpha'=\alpha'(\alpha)};\qquad \widetilde\gamma(\alpha) = \widetilde\gamma'(\alpha').
\end{equation}

\noindent
Using these equations and Eq. (\ref{Modified_NSVZ}), it is possible to relate the coupling constants in $\overline{\mbox{DR}}$ and NSVZ schemes \cite{Jack:1996vg},

\begin{equation}\label{DR_NSVZ_Renormalization}
\alpha' = \alpha + \frac{\alpha^3 N_f}{4\pi^2} + O(\alpha^4).
\end{equation}

\noindent
The $\overline{\mbox{DR}}$-scheme is defined by the boundary conditions (\ref{DR_Boundary_Conditions}). Therefore, for the NSVZ-scheme we obtain

\begin{equation}\label{NSVZ_Scheme}
\lim\limits_{\varepsilon\to\infty} \alpha_0(\alpha',\varepsilon, x_0=0)=\alpha' - \frac{\alpha'{}^3 N_f}{4\pi^2} + O(\alpha'{}^4);\qquad \lim\limits_{\varepsilon\to\infty} Z'(\alpha',\varepsilon, x_0=0) =1.
\end{equation}

\noindent
These conditions are the three-loop DRED analogs of Eqs. (\ref{NSVZ_HD_Scheme}). They differ from Eq. (\ref{DR_Boundary_Conditions}), because the NSVZ scheme is related with the $\overline{\mbox{DR}}$ scheme by the finite renormalization (\ref{DR_NSVZ_Renormalization}).\footnote{For $z(\alpha)\ne 1$ it is possible to construct boundary conditions giving a class of the NSVZ schemes. Then the right hand sides of the equations in (\ref{NSVZ_Scheme}) will be different. For example, the boundary conditions $Z_3(\alpha',\varepsilon\to \infty, x_0=0)=1$; $Z'(\alpha',\varepsilon\to \infty, x_0=0) = 1 +\alpha'/4\pi + O(\alpha'{}^2)$ give another NSVZ scheme.} The boundary conditions (\ref{NSVZ_Scheme}) give the following values of the finite constants in Eqs. (\ref{Three-Loop_Alpha}) and (\ref{Two-Loop_Z}):

\begin{equation}
g_1= g_2 = 0;\qquad b_1 =0;\qquad b_2 = -\frac{1}{4}.
\end{equation}

\noindent
It is easy to verify explicitly that in this case the NSVZ relation for the RG functions defined in terms of the renormalized coupling constant is really valid.

It is necessary to note that the NSVZ scheme defined by the prescription (\ref{NSVZ_Scheme}) is different from the NSVZ scheme which is obtained by using the higher derivative regularization and the boundary conditions (\ref{Boundary_Conditions}). Really, according to \cite{Kataev:2013csa}, if the NSVZ scheme is constructed with the higher derivative regularization, the RG functions are

\begin{eqnarray}
&& \widetilde\beta_{\mbox{\scriptsize HD}}(\alpha_{\mbox{\scriptsize HD}}) = \frac{\alpha_{\mbox{\scriptsize HD}}^2 N_f}{\pi}\Big(1 + \frac{\alpha_{\mbox{\scriptsize HD}}}{\pi} - \frac{\alpha_{\mbox{\scriptsize HD}}^2}{2\pi^2} - \frac{\alpha_{\mbox{\scriptsize HD}}^2 N_f}{\pi^2}\Big(1+\sum\limits_{I=1}^n c_I \ln a_I \Big) + O(\alpha_{\mbox{\scriptsize HD}}^3) \Big);\nonumber\\
&& \widetilde\gamma_{\mbox{\scriptsize HD}}(\alpha_{\mbox{\scriptsize HD}}) = - \frac{\alpha_{\mbox{\scriptsize HD}}}{\pi} + \frac{\alpha_{\mbox{\scriptsize HD}}^2}{2\pi^2} + \frac{\alpha_{\mbox{\scriptsize HD}}^2 N_f}{\pi^2} \Big(1+\sum\limits_{I=1}^n c_I \ln a_I \Big) + O(\alpha_{\mbox{\scriptsize HD}}^3).
\end{eqnarray}

\noindent
Here the coefficients $a_I\equiv M_I/\Lambda$ are the ratios of the Pauli--Villars masses $M_I$ to the parameter $\Lambda$ in the higher derivative regularizing term. They are assumed to be independent of the coupling constant. The degrees of the Pauli--Villars determinants entering into the generating functional are $N_f c_I$. The notation for the theory regularized by higher derivatives is described in \cite{Kataev:2013csa} in details. From the other side, in the NSVZ scheme constructed with the dimensional reduction the RG functions have the form

\begin{eqnarray}
&& \widetilde\beta'(\alpha') = \frac{\alpha'{}^2 N_f}{\pi}\Big(1 + \frac{\alpha'}{\pi} - \frac{\alpha'{}^2}{2\pi^2}(1+N_f) + O(\alpha'{}^3)\Big);\nonumber\\
&& \widetilde\gamma'(\alpha') = -\frac{\alpha'}{\pi} + \frac{\alpha'{}^2}{2\pi^2}(1+N_f) + O(\alpha'{}^3).
\end{eqnarray}

\noindent
We see that in both schemes the NSVZ relation is valid, but the renormalization group functions do not coincide. Therefore, these two NSVZ schemes can be related by a non-trivial finite renormalization (\ref{Finite_Renormalization}). From (\ref{RG_Transformation}) it is possible to find

\begin{eqnarray}
&& z(\alpha') = 1 - z_1 \frac{\alpha'}{\pi} + O(\alpha'{}^2);\qquad\qquad (\phi_R)_{\mbox{\scriptsize HD}} = z(\alpha')^{-1/2} \phi_R';\nonumber\\
&& \frac{1}{\alpha_{\mbox{\scriptsize HD}}} = \frac{1}{\alpha'} - \Big(\frac{1}{2} +\sum\limits_{I=1}^n c_I \ln a_I + z_1\Big) \frac{N_f}{\pi} - z_1 \frac{\alpha' N_f}{\pi^2} + O(\alpha'{}^2),
\end{eqnarray}

\noindent
where $\alpha_{\mbox{\scriptsize HD}}$ and $\alpha'$ are the coupling constants corresponding to the NSVZ schemes obtained with the higher derivative regularization and the dimensional reduction, respectively, and $z_1$ is an undefined finite constant. For example, one can set $z_1=0$ (or, equivalently, $z(\alpha)=1$ in the considered approximation), so that the finite renormalization does not change the matter superfields. The arbitrariness of choosing the constant $z_1$ follows from the arbitrariness of choosing the normalization point $\mu$.

\section{Conclusion}
\hspace*{\parindent}

In this paper in the three-loop approximation we find the relation between the RG functions of ${\cal N}=1$ SQED with $N_f$ flavors, regularized by the dimensional reduction, by the method similar to the one which has allowed constructing the all-loop NSVZ scheme for the theory regularized by higher derivatives. Because within the dimensional technique the limit of the vanishing external momentum is not well-defined, the loop integrals are not integrals of double total derivatives as in the latter case. However, using the analogous (but more complicated) structures, constructed in \cite{Aleshin:2015qqc}, we demonstrate that it is possible to relate the three-loop $\beta$-function to the two-loop anomalous dimension. Unlike the case of using higher derivatives, in the theory regularized by the dimensional reduction the RG functions defined in terms of the bare coupling constant do not satisfy the NSVZ relation. Due to the existence of an additional term in this relation, which has been calculated from the first principles, the RG functions (defined in terms of the renormalized coupling constant) in the $\overline{\mbox{DR}}$-scheme also do not satisfy the NSVZ relation. However, in the considered approximation it is possible to impose boundary conditions to the renormalization constants giving the NSVZ scheme with the dimensional reduction for the RG functions defined in terms of the renormalized coupling constant. These boundary conditions are similar to the ones obtained with the higher derivative regularization in all orders \cite{Kataev:2013eta}, but more complicated due to necessity of making the finite renormalization. Unlike the higher derivative regularization, we do not know, if it is possible to construct them in an arbitrary order, because the structure of the loop integrals in higher orders is not quite clear. Moreover, in higher loops the inconsistency of the dimensional reduction can be essential. It is important that the NSVZ schemes which have been constructed with the dimensional reduction and with the higher derivative regularization are different. They can be related by a finite renormalization, which is constructed in this paper in the lowest approximation.

\bigskip

\section*{Acknowledgments}
\hspace*{\parindent}

The work of A.K. was supported in part by the Russian Science Foundation Grant No. 14-22-00161. The work of K.S. was supported by the Russian Foundation for Basic Research, grant No. 14-01-00695.

\end{document}